\documentclass[aps,  showpacs]{revtex4}

\begin{document}
\title{ Unified field theories and Einstein}
\author{S. C. Tiwari \\
Department of Physics, Institute of Science,  Banaras Hindu University,  and Institute of Natural Philosophy, \\
Varanasi 221005, India }
\begin{abstract}
Einstein's contribution to relativity is reviewed. 
It is pointed out that Weyl gave first unified theory of gravitation and electromagnetism and it was different 
than the five dimensional theory of Kaluza. Einstein began his work on unification in 1925 that continued all through the rest of his life. 
A discussion is presented on the recent advances in Weyl theory, and also on the unification approach in which space-time is believed 
not to be fundamental.
The significance of the gravitational waves observed recently seems to indicate a new paradigm of unification. It is suggested that
the nature of time is the most fundamental issue for a break-through in the unification quest..
\end{abstract}
\maketitle

\section{Introduction}
Metaphysical concepts have an important role in science; however relativity and 
quantum revolutions have introduced avoidable weirdness and mystery in physics. News 
media and popular scientific literature alone cannot be blamed for this. Of late, the 
science journals and scientists themselves have been promoting cult figures and 
mysticism. UK astrophysicist Barrow \cite{1} says that unlike the scientific celebrities 
Newton and Darwin, Einstein has become an icon as 'Einstein restored faith in the 
unintelligibility of science.'  Does incomprehensible science not contradict the basic tenet 
of science? It is true that there are many scientists who subscribe to Barrow's view, but 
then the whole raison d'etre of scientific pursuit would become questionable. Synge 
writing on general relativity \cite{2} states:``The name is repellent. Relativity? I have never 
been able to understand what that word means in this connection. I used to think that this 
was my fault, some flaw in my intelligence, but it is now apparent that nobody ever 
understood it, probably not even Einstein himself. So let it go. What is before us is 
Einstein’s theory of gravitation''. It is not that Synge's is an isolated skepticism expressed 
in 1966, the foundations of general relativity have been criticized since 1917 beginning 
with the objections raised by E. Kretschmann \cite{3}. In quantum mechanics, Copenhagen 
interpretation defies scientific logic, and many physicists approvingly quote Feynman's 
verdict that no one really understands quantum mechanics, see a critique in \cite{4}. 
An individual is free to idolize Einstein or Bohr; however the purpose of science 
would be served better if the charismatic spell is kept out of the scientific discussions. 
Scientific spirit demands us to take cognizance of the struggle of the finest minds that 
laid the ground work for the creation of relativity and quantum theory, and to recognize 
the importance of constructive criticisms and dissent on the foundational issues. I vividly 
recall the inaugural lecture of V.V. Narlikar at the Einstein Centenary Symposium \cite{5} in which he remarked, 'Einstein was 
a rebel and a creator...'  Let us try to understand 
Einstein's role in the creation of relativity, and his place in the development of classical 
unified theories. 

\section{Creation of Relativity}

It is now well established \cite{5, 6} that Michelson-Morley experiment was known to 
Einstein before his 1905 relativity paper appeared and that since 1895 he had been 
occupied with the problem of the propagation of light if an observer followed the light 
beam with the same speed. Poincare and Lorentz made seminal contributions to the 
relativity theory, and in spite of the fact that Einstein did not cite their work even the title 
of this paper is inspired by the titles of the papers of Lorentz and Poincare. Whittaker in 
his treatise \cite{7} calls relativity – a theory of Poincare and Lorentz. However a careful 
review \cite{8} shows that the final decisive step in the creation of special relativity does 
belong to Einstein. It is evident that special relativity did not appear suddenly from no-where as a single man creation. 

There have been extensive critiques/elaborations on the 
special relativity, notable among them are due to H. Dingle, M. Bunge, and H. 
Reichenbach, see \cite{8} for references. In 1908 H. Minkowski introduced four dimensional 
space-time continuum which to Einstein appeared as 'superfluous learnedness'. 
Symmetry (rotational and translational) of space-time geometry and Lorentz 
transformations, and the treatment of the Maxwell field equations under the wider group 
of conformal transformations entered later in 1910 with the work of E. Cunningham and 
H. Bateman. Though there exist learned discussions on the paradoxes in special relativity, 
I arrived at a startling conclusion: Einstein mistook measurement convention of Newton's 
common time as Newtonian absolute time, and relativity does not address the question of the 
absolute time. I believe this misunderstanding has led to counter-intuitiveness in 
relativity; I refer to a detailed discussion in \cite{8} and a short essay posted on the arxiv \cite{9}. 

Historical survey by Whittaker \cite{7} and a scholarly review on the eight decades of 
dispute over the nature of general relativity \cite{3} show that this theory grew out of the 
important contributions of many scientists and its foundations are still not secure. 
Poincare argued that gravity must propagate with the speed of light and that Newton's 
law of gravitation had to be modified. In 1907 Planck noticed the significance of the equality of the inertial mass 
and gravitational mass (the Eoetvoes experiment) suggesting 
that 'all energy must gravitate'; six months later Einstein enunciated his preliminary 
version of the equivalence principle. Max Abraham as a powerful opponent and the 
mathematician Marcel Grossmann as a friendly collaborator were the key figures in 
shaping Einstein's thoughts on general relativity. In the 'Entwurf' theory of 1913 the 
physical and the mathematical parts were written by Einstein and Grossmann 
respectively. In fact, Bateman first saw the importance of the tensor calculus of Ricci and 
Levi-Civita. In a series of four papers during November-December of 1915, Einstein 
arrived at the final form of the gravitational field equations. 

There have been two crucial issues related with general relativity: (i) the role of 
general covariance, and (ii) the priority issue i.e. Hilbert or Einstein. In the Entwurf 
theory the Ricci tensor appears as the gravitational field tensor in the non-flat geometry. 
The role of general coordinate transformations and the Newtonian limit of the field 
equations became controversial. Stachel \cite{10} remarks that, ``He remained wedded to the 
Einstein-Grossmann field equations and kept trying to find better and better arguments in 
their favor; in particular, arguments for their uniqueness and for their invariance under 
the maximum invariance group compatible with avoiding the 'hole argument’.'' The hole 
argument in essence rejects general covariance. In June-July 1915 Einstein was in 
Goettingen at Hilbert's place and had intense correspondence with him until he reached 
the final field equations returning to general covariance. Einstein had received a draft of 
Hilbert's article before 18 November, 1915 that apparently contained the correct form of 
the field equations derived from an action principle. On the other hand, it has been 
pointed out \cite{10} that Einstein suspected an attempt at plagiarism by Hilbert which led to 
the strained relationship between them. Though Stachel settles the issue in favor of 
Einstein \cite{10}, see also Mehra \cite{11} it is obvious that Hilbert's pivotal role in the 
mathematical foundation of general relativity cannot be ignored. 

In 1918 Einstein elucidated general relativity in terms of three principles: principle of relativity, equivalence 
principle and Mach's principle. Since all the three 
principles have found varied interpretations and strong criticisms in the literature \cite{3}, I 
will make few remarks concerning the Kretschmann's objection against general covariance and its constructive alternative in the 
form of 'geometry-free physics' initiated 
by F. Kottler in 1922. Kretschmann's main argument is that all physical observations 
depend on purely topological relations between the objects in the space-time, and 
therefore no coordinate system is privileged. This implies that any theory could be so 
formulated mathematically that it is covariant under any group of coordinate 
transformations: general covariance is physically vacuous \cite{3}. General covariance in 
modern mathematical literature is called diffeomorphism: the group of transformations is 
differentiable point transformation on a differentiable manifold (just as isomorphism is in 
the vector space). Diffeo(4) in space-time manifold of general relativity takes into 
account accelerated or noninertial frames of reference. Post has drawn attention to what 
he calls KCD (Kottler, Cartan and van Dantzig) procedure \cite{12} for the electromagnetism. 
In a recent monograph Post articulated quantum cohomology as an alternative to quantum 
field theoretic unification schemes \cite{13}. In the light of current developments in the 
superstring theory, it is worth quoting Post on metric-free physics \cite{13}: ``Witten calls 
attention to metric-free aspects of these developments. This 'new' metric independence 
has, so far, not shown an awareness of the earlier metric-independent work of the 
Twenties and Thirties (compare index references to metric-independence). Yet, these 
recent reports can be taken as an encouragement to support Witten in his call for a metric-free (extended) 
principle of general covariance for space-time physics, because this 
extended covariance permits the one and only invariant reconciliation between quantum 
principles and general theory of relativity''. In his 1979 article \cite{12} Post has rightly 
stressed that KCD procedure is physically more encompassing and that it was 
independently formulated by ``three experts in the theory of differential and integral 
invariants''. 

\section{Unified theories}

Gustav Mie's theory was a precursor to the first attempt at a unified description of 
gravitation and electromagnetism by Hilbert. The nonlinear theory of Mie was aimed at 
explaining electron and matter from the fields, and the origin of variational principle for a 
world-function is due to him that became a powerful method in the hands of Hilbert and 
Weyl \cite{7, 14}. An importat advancement in mathematics was the notion of infinitesimal parallel transport 
of a vector by Levi-Civita in 1917. Weyl \cite{14} approached foundations 
of Riemannian geometry from this point of view, and was led to a pure infinitesimal 
geometry in which not only the direction of a vector under parallel transport from one 
point to another is changed but also its magnitude or length. Recall that in flat Euclidean 
geometry vectors at arbitrarily distant points can be compared as vector transference from 
one point to another is possible without any change in its direction and length. In the 
infinitesimal geometry of Riemann parallel transport of a vector from one point to 
another distant point rotates its direction. Weyl argued that there remained an element of 
finite geometry since the lengths of the vectors at distant points could still be compared. 
If the length of a vector is arbitrary upto a calibration function, the wider group of 
transformations makes it possible to introduce a distance curvature determined by a linear 
one-form (or vector potential) arising out of calibration transformation. Later Weyl called 
it gauge invariance translating the original Eichin-varianz. Weyl was not content with 
this geometry as being just a mathematical curiosity: he interpreted the distance curvature 
as physical electromagnetic field tensor and the generalized geometry as a unified theory 
of gravitation and electromagnetism. Hendry \cite{15} gives an account of Einstein-Weyl 
correspondence on this theory when in March 1918  Weyl sent his work to Einstein. At 
first Einstein was greatly impressed but soon raised physical objections: the change in the 
length of a vector should show up in the atomic spectra as the period of clocks would 
change over a lapse of time, however no such observation exists. In May 1918 Weyl 
wrote to Einstein that he was reluctant 'to accuse God of mathematical inconsistency' to 
which Einstein responded saying, 'It seemed to him just as bad to accuse God of a 
theoretical physics that did not do justice to human observations'. Weyl's metaphysical 
arguments on mathematical laws of nature appealed to Eddington, but he rejected Weyl's 
theory on both physical and mathematical grounds and proposed a generalized theory in 
which the assumption of the gauge invariance of zero length of a vector was removed \cite{16} 

Today we know that Weyl abandoned his original gauge theory in favor of phase 
transformations in Schroedinger and Dirac quantum theory, and it is this version that 
represents modern gauge field theories. Historically Fock in 1926 gave the first treatment of U(1) gauge 
transformation, and in 1927 F. London proposed quantum mechanical 
interpretation of Weyl's gauge theory. However two papers written in 1929 by Weyl are 
considered as landmark in the development of modern gauge theories \cite{17}. The beauty of 
original Weyl geometry attracted Dirac to revive it in 1973 \cite{18}. I have attempted to 
show that change of the length of a vector under parallel transport in complex space 
could be affected in such a way that the quantum state space retains typical phase 
characteristics \cite{19}. In later years, for a while, Einstein also returned to this geometry in 
the search for a unified theory.

Th. Kaluza in 1921 introduced fifth dimension to the space-time and originated a 
new direction to dissolve the duality of gravitation and electricity for a unified picture of 
nature. Weyl's theory of 1918 was characterized by him as a 'surprisingly courageous 
attack' to the unification problem. In Kaluza's theory the fifth coordinate is a new 
parameter and it is assumed that the derivatives with respect to this coordinate vanish (i.e. 
the cylinder condition); three-index Christofell symbol is sought to be interpreted as 
electromagnetic field tensor. Kaluza recognized physical and epistemological difficulties 
in his theory, mentioned the importance of quantum theory, and concluded that, ``If it 
would be proven some day that there exists more behind the presumed relations than 
merely meaningless formalism, then this would certainly imply a triumph for Einstein's 
general theory of relativity whose appropriate application to the five-dimensional world 
is at issue''. O. Klein in 1926 inspired by the new quantum theory (i.e. the wave 
mechanics) of de Broglie and Schroedinger re-interpreted Kaluza's theory treating the fifth 
dimension 'purely harmonic with a definite period related to the Planck constant'. Klein 
made an interesting insightful comment: the observed motion of a particle could be 
considered as the projection of wave motion in five dimension on the four dimensional 
space-time. In the current literature it is essentially due to the superstrings that Kaluza- 
Klein theory is widely known \cite{20}. Note that Weyl's was the first unified theory, and it 
was entirely different than the higher dimensional theory of Kaluza. 

It is curious that the papers of both Weyl and Kaluza were communicated by 
Einstein; he had serious doubts on these theories; and he himself was a late entrant to the 
quest of unification. Just before his death, Einstein revised the appendix on non-symmetric field in his 
book \cite{21} with the remarks: ``The last step of the theory concerns 
the unification of the field concept, which is characterized by the transition to non-symmetric fields. 
The difficulty in the choice of the field laws has been fully overcome 
only in the last few months. The arguments essential for this are presented in detail in 
Appendix-II''. Beginning with non-symmetric metric tensor field in 1925 Einstein 
explored many ideas for unifying gravitation and electromagnetism all through the rest 
of his life, and on several occasions he thought he had reached the goal but soon found 
them unsatisfactory. I think quite succinct opinion on this phase of Einstein's struggle is 
that of V.V. Narlikar \cite{5}: ``The creation of the complete theory or the total field theory 
was the first priority programme of Einstein from 1925 to 1954 before any application to 
cosmology or to situations demanding a quantum theory of gravity could be thought of. 
By total field Einstein originally meant a generalized field including both gravitational 
and electromagnetic fields and their interactions. Later he imposed a requirement 
whereby the theory included the Planck's constant h in an unforced manner..... Einstein, 
in his last paper on the subject, admitted that perhaps the concept of field was inadequate 
for the unified theory which he was seeking''. 

Eddington \cite{16}, Einstein \cite{21}, and later Schroedinger \cite{22} investigated purely 
affine theory, and for generalized theory dropped the assumption of symmetric affine 
connection. Recently I became aware of a revealing correspondence between Cartan and 
Einstein \cite{23} on the origin of absolute parallelism. I quote from Cartan's 8 May, 1929 
letter, 'In my terminology, spaces with a Euclidean connection allow of a curvature and a 
torsion: in the spaces where parallelism is defined in the Levi-Civita way, the torsion is 
zero; in the spaces where parallelism is absolute (fem parallelismus) the curvature is zero, 
thus these are spaces without curvature and with torsion'. Cartan pointed out that 
Einstein's new theory of generalized relativity introducing fem parallelismus in his 1928 
papers was a special case of Cartan's 1922 paper published in Comptes Rendus. Cartan 
also drew attention of Einstein to the discussion on this issue with him at Hadamard's 
home in 1922. Einstein in his reply accepted the priority claim of Cartan, and admitted 
that he did not understand Cartan's explanations in 1922 (at Hamamard's home). At the 
end of the letter he wrote, ``Asking you to forgive my inadvertent plagiarism and to help 
me settle the matter satisfactorily to everyone's benefit''. For a recent extensive review on 
unified theories I refer to Goenner's article \cite{24}.
 
\section{Discussion and Conclusion}

Serious and inquisitive reader would do well to study the original literature for a balanced 
and accurate view on Einstein and his work. Some of the original papers are cited here, and others could be found in the 
reviews/books referred to here. In a letter of 2 May, 1920 \cite{25} Einstein wrote 
to Bohr, ``Not often in life was I so delightfully impressed already by the mere presence 
of somebody as by yours''. Contemporary generation of physicists/philosophers look 
Einstein with awe, and barring few, have failed to raise basic questions on the relativity 
revolution. Should modern physics be allowed to remain captive to Einstein’s charm and 
ensuing weird physics? Remember that Einstein, though admired Bohr, continued to 
sharpen his arguments against Copenhagen interpretation of quantum mechanics. An 
objective assessment on Einstein, not anti-Einsteinian, rather than eulogizing him as a 
superman would surely serve the science and future generations better. In my opinion, 
reflecting upon the development of special relativity (from 1895 to 1905) and general 
relativity (from 1907 to 1915), and perusal of historical literature and Einstein-Cartan 
correspondence, Einstein emerges as a slow learner with average mathematical abilities 
but possessing an uncanny trait for perseverance in attacking the most difficult problems 
in physics. It is his imagination power and the courage to break from the trodden path at 
crucial juncture which set him apart from others; however he needed a framework for a 
zig-saw puzzle assembled by others from different pieces to take lead in making a 
complete picture. In the case of unified field theories such a framework did not exist, and 
Einstein himself had to struggle to reach at the different strands of unity. Single-minded 
pursuit in this endeavor makes him a thought personified. 

Let us have a brief discussion on recent unification efforts most prominent of them being the superstring theory.
In spite of great expectations from the superstrings, I believe this theory cannot 
represent physical reality, and simple ideas which make drastic revision of the space-time 
structure of relativity would be needed \cite{26}. Atiyah et al \cite{27} celebrate 50 years of twistor theory
admitting that this theory could merely reformulate known physical theories so far, but suggest  
holomorphic string theory in twistor space hold promise for future progress in unification. This article rightly notes
impressive advances in mathematics inspired by twistors, however I think in view of the fact that
compact holomorphic curves in a complex 3-fold in the twistor space is fundamental and space-time is secondary
this approach may not represent physical reality. Initially Penrose \cite{28} set the objective to be the reformulation in an approach
where continuum is replaced by discrete structures, for example, spin network, and construct space from this. The combined
algebra for linear momentum and angular momentum is the algebra of twistors: twistor is a spinor in the 6 dimensional
pseudo-orthogonal group O(4,2). In \cite{28} the Editor remarks that, 'did Penrose put the space-time structure into his theory
or did he deduce it from the theory'. Reading \cite{27} we find that this question remains unanswered even now.

The fundamental problem may lie in the concept of time. I suggest Einstein's relativistic time is not physical; it is 
not that we will return to Newtonian absolute time and pre-1905 physics, the role of time 
as a profound embodiment of creation and manifestation of universe would play a crucial 
role in future development of physics \cite{29}. Even original Weyl geometry for unification has immense potential for
new physics in relation to quantum theory and general relativity; I refer to a review by Scholz \cite{30}. Note that Scholz's review
is incomplete: a generalization to Weyl-Kaehler space \cite{19} and new action principle generalizing Weyl-Dirac theory \cite{31}
also deserve attention \cite{32}.

The discovery of gravitational waves in 2015 at the LIGO detectors \cite{33} is likely to introduce new dimension to unification 
paradigm. Does this observation validate Einstein's general relativity? This is a complex question. Recall that though Einstein
introduced gravitational waves in 1916, he arrived at a result in 1936 that they do not exist in an unpublished paper written with N. Rosen, see 
Kennefick \cite{34}. In the later published version this result was altered, however one needs to explore the possibility for alternative
explanation of gravitational waves. In fact, an interesting approach seems to be to use Kerr-Schild form of the metric tensor 
in the flat space-time \cite{35}.

{\bf Acknowledgements}

I gratefully acknowledge correspondence with M. Atiyah, 
E.J. Post and E. Scholz. I thank Jim Andrakin for sending me the book on Cartan-Einstein correspondence.

\end{document}